\documentclass[pra,twocolumn,cuted,showpacs,preprintnumbers,amsmath,amssymb]{revtex4}

\usepackage{amsmath,amssymb}
\usepackage[boxruled]{algorithm2e}
\usepackage{algpseudocode}
\newcommand{\gsgs}{\Phi_{\mbox{\tiny HF}}}
\newcommand{\gs}{\cdot}

\newcommand{\up}[1]{a_{#1}^+}
\newcommand{\dn}[1]{a_{#1}}
\newcommand{\bra}[1]{\left< #1\right|}
\newcommand{\ket}[1]{\left|#1 \right>}

\newcommand\double[2]{\left<#1|#2\right>}
\newcommand\ddouble[2]{\left<#1||#2\right>}

\newcommand\braket[3]{\left<#1\left|#2\right|#3\right>}
\newcommand\hfbraket[1]{\braket{\gs}{#1}{\gs}}

\newcommand\emphasize[1]{{\underline{#1}}}
\newcommand\eqRef[1]{Eq.\ \ref{#1}}
\newcommand\secRef[1]{Sec.\ \ref{#1}}
\newcommand\algoRef[1]{Algo.\ \ref{#1}}

\begin{document}

\author{Xinle Liu}
\email{lxinle@sas.upenn.edu}
\affiliation{Dept.\ of Chemistry, University of Pennsylvania, Philadelphia, PA 19104}

\begin{abstract}
Second quantization has been widely used in quantum mechanics and quantum chemistry, which is trivial and error-prone for researchers.
Fortunately it is a good candidate for automatic evaluation with its simple, trivial and intrinsic iterative nature.
This article presents an automatic and generic tool for these kinds of evaluations.
We believe it could be a helpful tool for scientists,
and hope the idea of automatic evaluation could be generalized to other equation derivations in the future.
\end{abstract}

\title{A generic tool to evaluate second quantization}
\maketitle

\section{Introduction}
\label{sec:intro}
With development of modern computing powers, massive scale scientific computing has been feasible and providing new evidence and insights into understanding scientific problems.
Thus it is possible for computers to perform more high-levels of calculations which has been impossible in the past years, which could provide lots of opportunities and driving forces for science.
Second quantization has been a basic tool to evaluate expressions in quantum mechanics and mechanics,
with electron creation ($a_p^+$) and electron annihilation ($a_p$) operators, with basic identities \cite{szabo:ostlund}:
\begin{equation}
\label{eq:second_id_commute}
\left\{
\begin{array}{llll}
\{a_p, a_q\} &= 0\\
\{a_p^+, a_q^+\} &= 0\\
\{a_p^+, a_q\} &= \delta_{pq}
\end{array}
\right.
\end{equation}
In quantum chemistry, they could be evaluated at different levels with Configuration Interactions (CI),
together with different levels of interaction which is reflected in the number of the operators as well.
When evaluating one wavefunction interactions with another, there are 3 sources of operators,
one from the bra wavefunction ($\bra{\Psi}$),
one from the ket wavefunction ($\ket{\Psi}$),
and another from the Hamiltonian matrix element, which could be single electron operator, double electron operator, etc.
All together, there could be multiple operators involved in the evaluation, and takes time to evaluate.
With a simple example of double electron operator with two CI doubles wavefunction,
there are 4 operators from each of those 3 sources, leading to a total of 12 operators.
The evaluation of the final expression is trivial, time-consuming and error prone for researchers.
Fortunately, the evaluation rule is very simple according to identity \eqRef{eq:second_id_commute}, and two trivial terminal states with a vacuum state \cite{szabo:ostlund}:
\begin{equation}
\label{eq:second_id_empty}
\left\{
\begin{array}{llll}
a_p &|> &=0 \\
< | &a_p^+ &= 0\\
\end{array}
\right.
\end{equation}
where $p$ could be any molecular orbital.

\section{Algorithm}
{\bf Notation}. Throughout this article, ${ijklmn,\ abcd \mbox{ and } pqrs}$ denote
\textit{occupied}, \textit{virtual} or \textit{any} molecular orbitals respectively,
with the ground state wavefunction as a reference.
Also whenever an index appears twice in an expression, it implies a sum over this index.

In this section, we show how the program works for evaluating expressions with lots of operator involved.
First we define a few identities that could be used to simplify expressions, second the expression go through iterations with those rules until no more operations are possible,
in the end, terms could be combined according to their symmetric and antisymmetric properties.

\subsection{Identities}
There are a few identities that could be used to simplify expressions with second quantization,
starting with the basic identities defined in \eqRef{eq:second_id_commute} and \eqRef{eq:second_id_empty} in \secRef{sec:intro},
it is pretty straightforward to show:
\begin{equation}
\label{eq:second_id_ket}
\left\{
\begin{array}{lllll}
a_i^+ &\ket{\gs} &= 0\\
a_a &\ket{\gs} & =0 \\
\end{array}
\right.
\end{equation}
where $\ket{\gs}$ is a shorthand notation for the ground state wavefunction $\ket{\gsgs}$.
With this equality, operators $a_i^+$ and $a_a$ should move towards the right hand side of expressions.

Similarly,
\begin{equation}
\label{eq:second_id_bra}
\left\{
\begin{array}{llllll}
\bra{\gs} & a_i &= 0\\
\bra{\gs} & a_a^+& =0 \\
\end{array}
\right.
\end{equation}
implies that operators $a_i$ and $a_a^+$ should move towards the left hand side of expressions.
Note that \eqRef{eq:second_id_bra} is simply the Hermitian Conjugate of \eqRef{eq:second_id_ket}.

There are other identities and definitions that could be used for simplification, which depends on the symmetry of Hamiltonian matrix elements,
for single electron Hamiltonian it reads as:
\begin{equation}
\label{eq:symmetry_o1}
\begin{array}{llll}
h_{pq} = h_{qp}\\
\end{array}
\end{equation}
and for double electron Hamiltonian, it follows:
\begin{equation}
\label{eq:symmetry_o2}
\left\{
\begin{array}{lllllll}
\double{pq}{rs} &= \double{rs}{pq} &= \double{qp}{sr} & = \double{sr}{qp} \\
\ddouble{pq}{rs} &\equiv \double{pq}{rs} & -\double{pq}{sr}\\
\end{array}
\right.
\end{equation}

\subsection{Iterations}
\label{subsec:iterations}
Next, depending on the iterative nature of evaluating those operators,
we present the flowchart for a automatic and generic tool in this article.

The very first step is the iteration phase, for each operator, to simplify the expression, operator should go along a certain directions.
With $a_i^+$ as an example, since $a_i^+ |\gs>$ is $0$, and expressions with this pattern could be removed from the iterations,
$a_i^+$ is supposed to swap position with its neighbors and go along to the right hand side.
Similarly, $a_a |\gs>= 0$ means that $a_a$ should go along to the right as well.
On the other hand, $<\gs| a_i = 0$ and $<\gs| a_a^+ = 0$ implies $a_i$ and $a_a^+$ should go to the left hand side.
As long as there are $a_i$, $a_i^+$, $a_a$ and $a_a^+$ operators in the expression, the iteration phase has not finished yet.
\algoRef{algo:one_step_iteration} shows the basic idea and how it is implemented for \textit{one} single iteration.

\begin{algorithm}[H]
  \KwData{A single term to evaluate $t$.}
  \KwResult{A list of terms with one step iteration: could be $0, 1$ or $2$ terms.}

  \eIf{$\exists o \in t$ that is supposed to move to the \emphasize{right}}{
    Find out the first $o$ from the \emphasize{right} hand side;

    \eIf{$o$ is the first operator from the \emphasize{right}}{
      return []; \Comment{$0$ term;}
    }{
      return terms from swapping order of $o$ and the operator on the \emphasize{right} $o_r$ according to \eqRef{eq:second_id_commute};

      \Comment{$1$ or $2$ terms depending on creation/ annihilation types and whether $\delta$ term is $0$;}
    }
  }{
    \eIf{$\exists o \in t$ that is supposed to move to the \emphasize{left}}{
      Find out the first $o$ from the \emphasize{left} hand side;

      \eIf{$o$ is the first operator from the \emphasize{left}}{
	return []; \Comment{$0$ term;}
      }{
	return terms from swapping order of $o$ and the operator on the \emphasize{left} $o_l$ according to \eqRef{eq:second_id_commute};

	\Comment{$1$ or $2$ terms depending on creation/ annihilation types and whether $\delta$ term is $0$;}
      }
    }{
      return $t$; \Comment{No operation;}
    }
  }
\caption{One step iteration for expressions in second quantization.}
\label{algo:one_step_iteration}
\end{algorithm}
Note that Iteration Phase won't finish until all terms are unchanged from the previous iteration.

\subsection{Evaluation}
\label{subsec:evaluation}
With the iteration phase done in \secRef{subsec:iterations}
Second step is the sort phase, which is optional and auxiliary to make it convenient for the combination phase of same terms.
Next comes the merge or combination phase, which is mainly to merge terms that are equivalent, e.g. $\double{pq}{rs}$ and $\double{qp}{sr}$,
while it is able to remove any terms that cancel each other as well.
The very last step is to simplify coefficients, which comes from $\delta$ functions for orbitals.
With this step, one can see clearly the anti-symmetric pattern for wavefunctions.

Here is a flowchart of the algorithm:

\begin{algorithm}[H]
  \KwData{A high level expression to evaluate $t$.}
  \KwResult{A list of terms down to single and double electron interactions.}

  \Comment{Iteration phase;}\\
  terms = [$t$]

  \While{True}{
    new\_terms = []

    \ForEach{term $\in$ terms}{
      new\_terms.append(evaluate(term));

      \Comment{With one-step iteration defined in \algoRef{algo:one_step_iteration};}
    }

    \If{terms $==$ new\_terms}{
	break;
    }
  }

  \Comment{Sorting phase;}\\
  Sort terms according to their string representation;

  \Comment{Merge phase;}\\
  Merge terms that are equal;

  \Comment{Coefficient substitution;}\\
  Remove unnecessary $\delta$ terms and its substitution in other parts of expressions;
 \caption{Flowchart to evaluate expressions in second quantization.}
\end{algorithm}


\section{Results}
\label{sec:results}
In this section, we show the results of a few applications of this algorithm, with a few typical expressions used in quantum chemistry.
Firstly it is the canonical evaluation for neutral molecules,
CI Singles ($\ket{\Psi_i^a}$),
CI Doubles ($\ket{\Psi_{ij}^{ab}}$) with
single electron Hamiltonian $h_1= h_{pq} \up{p} \dn{q}$ and
double electron Hamiltonian $h_2 = \up{p} \up{q} \dn{s} \dn{r} \double{pq}{rs}$,
which could be used to verify its validity easily.
More importantly, it could be generalized to evaluate \textit{any} interactions,
not only traditional neutral molecules,
but also cations with losing an electron from occupied orbital
as well as anions with an extra electron in the virtual orbital, etc.

\subsection{Singles}
\label{subsec:CIS}
CI Singles ($\ket{\Phi_i^a}$) is the simplest, least expensive and most widely used states to simulate molecular excited states, on top of ground states ($\ket{\gsgs}$), we'll start from here.

For single electron Hamiltonian, it reads as the following:
\begin{equation}
\label{eq:CIS_single}
\begin{array}{lllllll}
    &\braket{\Psi_{\mbox{{\tiny single}}}}{\hat{O}_1}{\Psi_{\mbox{{\tiny single}}}} = \\
    & + t_j^bt_i^a
    \hfbraket{a_j^+ a_b a_p^+ a_q a_a^+ a_i} h_{pq} \\
    &=
    - t_j^at_i^a h_{ij}
    + t_i^bt_i^a h_{ab}
    + t_i^at_i^a h_{mm}
\end{array}
\end{equation}

For double electron Hamiltonian, it follows:
\begin{equation}
\label{eq:CIS_double}
\begin{array}{lllllll}
    &\braket{\Psi_{\mbox{{\tiny single}}}}{\hat{O}_2}{\Psi_{\mbox{{\tiny single}}}} = \\
    &+0.5 t_j^bt_i^a  \hfbraket{a_j^+ a_b a_p^+ a_q^+ a_s a_r a_a^+ a_i} \double{pq}{rs} &=\\
    &- t_j^bt_i^a    \ddouble{aj}{bi}
     - t_j^at_i^a \ddouble{im}{jm}\\
    &+ t_i^bt_i^a \ddouble{am}{bm}
     + 0.5 t_i^at_i^a \ddouble{mn}{mn}\\
\end{array}
\end{equation}
It's easy to show that \eqRef{eq:CIS_single} and \eqRef{eq:CIS_double} together recovers the canonical CIS energies.

\subsection{Doubles}
\label{subsec:CID}
Though CI Singles approximation is normally enough for molecular excited states \cite{mhg:1994:cpl_cisd},
it is barely enough for Charge Transfer states,
as Subotnik \textit{et al} shows in a series of research papers \cite{
subotnik:2011:bias,
xinle:jcp_oocis:2012,
xinle:jcp_voocis:2013,
xinle:jctc_voocis:2014,
xinle:jpca_voocis:2015}.
With this in mind, there have been many ways to make corrections with CI singles,
various perturbative and varational approaches have been developed over the years,
with exmaples of CIS(D) \cite{mhg:1994:cpl_cisd}, Coupled Cluser methods\cite{bartlett:1982:ccsdt},
OO-CIS \cite{xinle:jcp_oocis:2012},
VOA-CIS \cite{ xinle:jcp_voocis:2013, xinle:jctc_voocis:2014}.
In all those approaches at least double excitations,
i.e. CI Doubles, relative to the HF ground state are involved.
Following equations show the results for CI doubles with single electron Hamiltonian:
\begin{equation}
\begin{array}{lllllll}
    &\braket{\Psi_{\mbox{{\tiny double}}}}{\hat{O}_1}{\Psi_{\mbox{{\tiny double}}}}= \\
    &+ t_{kl}^{cd}t_{ij}^{ab} \hfbraket{a_l^+ a_k^+ a_d a_c a_p^+ a_q a_a^+ a_b^+ a_i a_j} h_{pq} &=\\
    &+ t_{kl}^{cd}t_{ij}^{ab} a_l^+ a_k^+ a_d a_c a_p^+ a_q a_a^+ a_b^+ a_i a_j h_{pq} &=\\
    &
- t_{il}^{ab}t_{ij}^{ab}    h_{jl}
+ t_{ki}^{ab}t_{ij}^{ab}    h_{jk}\\&
+ t_{jl}^{ab}t_{ij}^{ab}    h_{il}
- t_{kj}^{ab}t_{ij}^{ab}    h_{ik}\\&
+ t_{ij}^{ad}t_{ij}^{ab}    h_{bd}
- t_{ji}^{ad}t_{ij}^{ab}    h_{bd}\\&
+ t_{il}^{ba}t_{ij}^{ab}    h_{jl}
- t_{ki}^{ba}t_{ij}^{ab}    h_{jk}\\&
- t_{jl}^{ba}t_{ij}^{ab}    h_{il}
+ t_{kj}^{ba}t_{ij}^{ab}    h_{ik}\\&
- t_{ij}^{ca}t_{ij}^{ab}    h_{bc}
+ t_{ji}^{ca}t_{ij}^{ab}    h_{bc}\\&
- t_{ij}^{bd}t_{ij}^{ab}    h_{ad}
+ t_{ji}^{bd}t_{ij}^{ab}    h_{ad}\\&
+ t_{ij}^{cb}t_{ij}^{ab}    h_{ac}
- t_{ji}^{cb}t_{ij}^{ab}    h_{ac}\\&
+ t_{ij}^{ab}t_{ij}^{ab} h_{kk}
- t_{ji}^{ab}t_{ij}^{ab} h_{kk}\\&
- t_{ij}^{ba}t_{ij}^{ab} h_{kk}
+ t_{ji}^{ba}t_{ij}^{ab} h_{kk}
\end{array}
\end{equation}

and with double electron Hamiltonian:
\begin{equation}
\begin{array}{lllllll}
  &\braket{\Psi_{\mbox{{\tiny double}}}}{\hat{O}_2}{\Psi_{\mbox{{\tiny double}}}}= \\
  & +0.5 t_{kl}^{cd}t_{ij}^{ab} \hfbraket{a_l^+ a_k^+ a_d a_c a_p^+ a_q^+ a_s a_r a_a^+ a_b^+ a_i a_j} \double{pq}{rs} &=\\&
+ t_{kl}^{ab}t_{ij}^{ab}    \ddouble{ij}{kl}
- t_{il}^{ad}t_{ij}^{ab}    \ddouble{bl}{dj}\\&
+ t_{ki}^{ad}t_{ij}^{ab}    \ddouble{bk}{dj}
+ t_{jl}^{ad}t_{ij}^{ab}    \ddouble{bl}{di}\\&
- t_{kj}^{ad}t_{ij}^{ab}    \ddouble{bk}{di}
- t_{kl}^{ba}t_{ij}^{ab}    \ddouble{ij}{kl}\\&
+ t_{il}^{ca}t_{ij}^{ab}    \ddouble{bl}{cj}
- t_{ki}^{ca}t_{ij}^{ab}    \ddouble{bk}{cj}\\&
- t_{jl}^{ca}t_{ij}^{ab}    \ddouble{bl}{ci}
+ t_{kj}^{ca}t_{ij}^{ab}    \ddouble{bk}{ci}\\&
+ t_{il}^{bd}t_{ij}^{ab}    \ddouble{al}{dj}
- t_{ki}^{bd}t_{ij}^{ab}    \ddouble{ak}{dj}\\&
- t_{jl}^{bd}t_{ij}^{ab}    \ddouble{al}{di}
+ t_{kj}^{bd}t_{ij}^{ab}    \ddouble{ak}{di}\\&
- t_{il}^{cb}t_{ij}^{ab}    \ddouble{al}{cj}
+ t_{ki}^{cb}t_{ij}^{ab}    \ddouble{ak}{cj}\\&
+ t_{jl}^{cb}t_{ij}^{ab}    \ddouble{al}{ci}
- t_{kj}^{cb}t_{ij}^{ab}    \ddouble{ak}{ci}\\&
+ t_{ij}^{cd}t_{ij}^{ab}    \ddouble{ab}{cd}
- t_{ji}^{cd}t_{ij}^{ab}    \ddouble{ab}{cd}\\&
- t_{il}^{ab}t_{ij}^{ab} \ddouble{jm}{lm}
+ t_{ki}^{ab}t_{ij}^{ab} \ddouble{jm}{km}\\&
+ t_{jl}^{ab}t_{ij}^{ab} \ddouble{im}{lm}
- t_{kj}^{ab}t_{ij}^{ab} \ddouble{im}{km}\\&
+ t_{ij}^{ad}t_{ij}^{ab} \ddouble{bm}{dm}
- t_{ji}^{ad}t_{ij}^{ab} \ddouble{bm}{dm}\\&
+ t_{il}^{ba}t_{ij}^{ab} \ddouble{jm}{lm}
- t_{ki}^{ba}t_{ij}^{ab} \ddouble{jm}{km}\\&
- t_{jl}^{ba}t_{ij}^{ab} \ddouble{im}{lm}
+ t_{kj}^{ba}t_{ij}^{ab} \ddouble{im}{km}\\&
- t_{ij}^{ca}t_{ij}^{ab} \ddouble{bm}{cm}
+ t_{ji}^{ca}t_{ij}^{ab} \ddouble{bm}{cm}\\&
- t_{ij}^{bd}t_{ij}^{ab} \ddouble{am}{dm}
+ t_{ji}^{bd}t_{ij}^{ab} \ddouble{am}{dm}\\&
+ t_{ij}^{cb}t_{ij}^{ab} \ddouble{am}{cm}
- t_{ji}^{cb}t_{ij}^{ab} \ddouble{am}{cm}\\&
+0.5 t_{ij}^{ab}t_{ij}^{ab} \ddouble{mn}{mn}
-0.5 t_{ji}^{ab}t_{ij}^{ab} \ddouble{mn}{mn}\\&
-0.5 t_{ij}^{ba}t_{ij}^{ab} \ddouble{mn}{mn}
+0.5 t_{ji}^{ba}t_{ij}^{ab} \ddouble{mn}{mn}
\end{array}
\end{equation}
For CI Doubles, it shows the handiness of an automatic and generic tool presented in this article,
otherwise it is $\sim100$ pages of derivation manually!

\subsection{Anions}
\label{subsec:anions}
Though second quantization has been widely used for neutral molecules,
it could be generalized for
anions $\up{a} \ket{\gsgs}$
and
cations $\dn{i} \ket{\gsgs}$
as well,
with very straightforward physical meanings for the ending expressions.

For single electron Hamiltonian, it follows:
\begin{equation}
\label{eq:anions_single}
\begin{array}{lllllll}
    &t^b \braket{\gs}{\dn{b} \hat{O}_1 \up{a}}{\gs} t^a= \\
    &+ t^bt^a \hfbraket{a_b a_p^+ a_q a_a^+} h_{pq} &=\\
    &+ t^bt^a    h_{ab}
+ t^at^a  h_{mm}\\
\end{array}
\end{equation}
Here, the first term results from adding an extra electron in a virtual molecular orbitals, which is expected.

For double electron Hamiltonian, it reads:
\begin{equation}
\label{eq:anions_double}
\begin{array}{lllllll}
    &t^b \braket{\gs}{\dn{b} \hat{O}_2 \up{a}}{\gs} t^a= \\
    &+0.5 t^bt^a \hfbraket{a_b a_p^+ a_q^+ a_s a_r a_a^+} \double{pq}{rs}&=\\
    &+ t^bt^a \ddouble{am}{bm}
+0.5 t^at^a \ddouble{mn}{mn}\\
\end{array}
\end{equation}

\subsection{Cations}
\label{subsec:cations}
Similar to \secRef{subsec:anions}, we can generalize it to anions as well with a clear explanation for those terms.

For single electron Hamiltonian, it follows:
\begin{equation}
\label{eq:cations_single}
\begin{array}{lllllll}
    &t_j \braket{\gs}{\up{j} \hat{O}_1 \dn{i}}{\gs} t_i= \\
    &+ t_jt_i \hfbraket{a_j^+ a_p^+ a_q a_i} h_{pq}=\\
    &- t_jt_i    h_{ij}
+ t_it_i a_p^+ a_q h_{pq}\\
\end{array}
\end{equation}
Here, the first term comes from losing an occupied electron in an occupied molecular orbital.

For double electron Hamiltonian, it reads:
\begin{equation}
\label{eq:cations_double}
\begin{array}{lllllll}
    &t_j \braket{\gs}{\up{j} \hat{O}_2 \dn{i}}{\gs} t_i= \\
    &+0.5 t_jt_i \hfbraket{a_j^+ a_p^+ a_q^+ a_s a_r a_i} \double{pq}{rs}=\\
    &- t_jt_i \ddouble{im}{jm}
+0.5 t_it_i \ddouble{mn}{mn}\\
\end{array}
\end{equation}

Also note that by combining
\eqRef{eq:anions_single} and \eqRef{eq:anions_double},
or
\eqRef{eq:cations_single} and \eqRef{eq:cations_double},
we're able to recover Koopmans' theorem \cite{levine:quantum_book}.
With more electrons involved,
this generic tool has the \textit{full} flexibility and capability to generalize,
which would be much more helpful with those cases.

\section{Conclusion}
Second quantization is widely used in quantum mechanics and quantum chemistry, to the best of authors’ knowledge, however,
there is no existing tool to do these kinds of evaluation automatically.
We believe this could be a useful and handy tool for researchers and motivating new ideas  in this fields in the coming years.
Moreover, there are plenty of space for automatic evaluation of equations in quantum mechanics
\cite{schatz:quantum_book,sakurai:modern},
a straightforward example would be the commonly text-book example of perturbation theory to higher order terms,
similar patterns could be seen in many other equations as well.

\section{References}

\end{document}